\documentclass[a4paper,fleqn,usenatbib]{mnras}

\usepackage{newtxtext,newtxmath}

\usepackage[T1]{fontenc}
\usepackage{ae,aecompl}
\usepackage{threeparttable}
\usepackage{bm}

\usepackage{graphicx}	
\usepackage{amsmath}	
\usepackage{amssymb}	

\title[Evolution of orientations of DM haloes and CGs] {Cosmological evolution of orientations of cluster-sized dark matter haloes and their central galaxies in the Horizon-AGN simulation}

\author[Taizo Okabe et al.]{Taizo Okabe$^{1}$,\thanks{E-mail: taizo.okabe@utap.phys.s.u-tokyo.ac.jp}
Takahiro Nishimichi$^{2,3}$,
Masamune Oguri$^{1,2,4}$,
S{\'e}bastien Peirani$^{5,6}$,
\newauthor
Tetsu Kitayama$^{7}$,
Shin Sasaki$^{8}$, 
Yasushi Suto$^{1,4}$,
Christophe Pichon$^{6,9}$,
\newauthor
and Yohan Dubois$^{6}$
\\
$^{1}$Department of Physics, The University of Tokyo, 7-3-1 Hongo, Bunkyo-ku, Tokyo 113-0033, Japan\\
$^{2}$Kavli Institute for the Physics and Mathematics of the Universe (WPI), The University of Tokyo Institutes for Advanced Study, \\
The University of Tokyo, 5-1-5 Kashiwanoha, Kashiwa, Chiba 277-8583, Japan\\
$^{3}$Center for Gravitational Physics, Yukawa Institute for Theoretical Physics, Kyoto University, Kyoto 606-8502, Japan\\
$^{4}$Research Center for the Early Universe, School of Science, The University of Tokyo, 7-3-1 Hongo, Bunkyo-ku, Tokyo, 113-0033, Japan\\
$^{5}$Universit\'e C\^ote d'Azur, Observatoire de la C\^ote d'Azur, \\
CNRS, Laboratoire Lagrange, Bd de l’Observatoire, CS 34229, 06304 Nice Cedex 4, France \\
$^{6}$Institut d'Astrophysique de Paris (UMR 7095: CNRS \& UPMC), 98 bis Bd Arago, 75014 Paris, France \\
$^{7}$Department of Physics, Toho University, Funabashi, 2-2-1 Miyama, Funabashi, Chiba 274-8510, Japan\\
$^{8}$Department of Physics, Tokyo Metropolitan University, 1-1 Minami-Osawa, Hachioji, Tokyo 192-0397, Japan\\
$^{9}$Korea Institute for Advanced Study, 85 Hoegiro, Dongdaemun-gu, Seoul, 02455, Republic of Korea
}

\date{Accepted XXX. Received YYY; in original form ZZZ}

\pubyear{2017}

\begin{document}
\label{firstpage}
\pagerange{\pageref{firstpage}--\pageref{lastpage}}
\maketitle

\begin{abstract}
  It is known observationally that the major axes of galaxy clusters
  and their brightest cluster galaxies are roughly aligned with each
  other.  To understand the origin of the alignment, we identify 40
  cluster-sized dark matter (DM) haloes with masses higher than
  $5\times10^{13}~M_{\odot}$ and their central galaxies (CGs) at
  $z\approx 0$ in the Horizon-AGN cosmological hydrodynamical
  simulation. We trace the progenitors at 50
    different epochs between $0<z<5$.  We then fit their shapes and
    orientations with a triaxial ellipsoid model.  While the
  orientations of both DM haloes and CGs change significantly due to
  repeated mergers and mass accretions, their relative orientations
  are well aligned at each epoch even at high redshifts, $z>1$.  The
  alignment becomes tighter with cosmic time; the major axes of the
  CGs and their host DM haloes at present are aligned on average
  within $\sim 30^\circ$ in the three dimensional space and $\sim
  20^\circ$ in the projected plane.  The orientations of the major
  axes of DM haloes on average follow one of the eigen-vectors of
  the surrounding tidal field that corresponds to the {\it slowest
    collapsing} (or even stretching) mode, and the alignment with the
  tidal field also becomes tighter.  This implies that the
  orientations of CGs and DM haloes at the present epoch are
  largely imprinted in the primordial density field of the Universe,
  whereas strong dynamical interactions such as mergers are important
  to explain their mutual alignment at each epoch.
\end{abstract}




\section{Introduction} \label{sec:intro}
Observed shapes of galaxies and galaxy clusters are not spherical, but
rather approximated well by ellipsoids.
Their orientations defined by the position angles of the major axes
may indicate a preferred direction in their formation process that is supposed to reflect the initial
condition and/or the dynamical evolution.

There are numerous observational studies that have reported
statistical correlations of those orientations over various scales.
One of the most well-known results is the alignment between the
orientations of the brightest cluster galaxies (BCGs) and their host
clusters \cite[e.g.][]{1968PASP...80..252S, 1980MNRAS.191..325C,
  1982A&A...107..338B}.  For instance, \cite{1982A&A...107..338B}
  reported that merger axes of 39 galaxy clusters at
  $z<0.1$ and their BCGs are aligned with $\sim 30^{\circ}$ on
  average.  This result has been studied further and confirmed for
  wider samples at different redshifts
  \cite[e.g.][]{2008MNRAS.385.1511W, 2009AJ....138.1709P,
    2011ApJ...740...39H, 2016MNRAS.463..222H}.  

More recently, \cite{2017NatAs...1E.157W} measured the alignment
between orientations of clusters and their BCGs, and obtained a mean
value of about $\sim 30^{\circ}$ for 52 clusters.  Their most
important finding is that the alignment extends to $z>1.3$ with high
statistical significance.  While the alignments are ubiquitous
observationally, their physical origin is not well understood, and
remains to be explained theoretically.

Cosmological (dark matter only) $N$-body simulations have been used
for understanding the origin of those observed alignments.  For
instance, there have been many previous attempts to examine the
alignment among the major axes of dark matter (DM) haloes at different
scales \cite[e.g.][]{1991ApJ...369..287W, 1994MNRAS.268...79W,
  1998ApJ...502..141D, 2002ApJ...574..538J, 2008ApJ...675..146F}.  In
particular, \citet{2002ApJ...574..538J} introduced triaxial modeling
of dark matter haloes in the cold dark matter (CDM) model, and found
that major axes of iso-density surfaces at different density thresholds in the
same halo are roughly aligned.  \citet{2016PASJ...68...97S} further
examined the evolution of DM haloes, and found that shapes and
position angles of the inner regions change significantly over the
cosmic time relative to the outer region of the same cluster-sized
haloes. They also found that around $z=0$, the inner region of those
haloes become rounder than the outer region, and tend to be aligned
toward the orientation of the host DM halo (see their Figure 4).

Since $N$-body simulations do not include baryon physics,
BCGs cannot be defined in a straightforward manner. It is not
clear to what extent the orientation of the inner region of those
DM haloes can be regarded as a good proxy for that of BCGs.  Reliable
predictions concerning the alignment between the orientations of BCGs
and their hosting DM haloes require cosmological hydrodynamical
simulations that incorporate proper baryon physics including gas
cooling, star formation, and supernova/AGN feedback as well.
Several previous attempts \cite[e.g.][]{2014ApJ...791L..33D, 2015MNRAS.453..721V,
  2015MNRAS.453..469T, 2017MNRAS.472.1163C} have found that
major axes of BCGs and their host DM haloes are fairly well aligned,
although the result should depend on how to implement baryon physics in a reliable fashion.
  
We are carrying out systematic studies of the non-sphericity and
orientation of galaxy clusters using the Horizon-AGN simulation \citep{dubois14}, 
a state-of-the-art cosmological hydrodynamical simulation incorporating
proper baryon physics.  \citet{suto17} (Paper I)
focused on the projected-axis ratios of the stellar component,
X-ray-emitting gas, and DM in 40 cluster-sized simulated DM haloes
with $M_{\rm DM} >5\times10^{13}M_{\odot}$, and showed that even
shapes of DM haloes in the outer region of clusters are substantially affected
by baryon physics. Indeed, the projected axis ratios of the
simulated haloes become consistent with those derived from the
observed X-ray clusters only when the AGN feedback is included.

\cite{okabe18} (Paper II) computed the position angles of DM, gas, and
stellar mass distributions for those 40 haloes at the present epoch,
$z\approx0$ alone, and examined the statistics of their mutual
alignment.  In particular, Paper II examined the difference of the
position angles $\Delta\theta$ of various components relative to that
of the CG in the same halo, and found that the root mean
square of $\Delta\theta$ is less than $25^\circ$, indicating that they
are relatively well aligned with each other. While this conclusion is
consistent with both previous simulations and observational results,
the origin of the alignment is not yet clear.  In this paper, we
extend Paper II and attempt to explain the origin by considering the
evolution of the alignment.

Furthermore, there are many recent studies that report the alignment
of clusters over $\sim100$ $h^{-1}$Mpc scales, and also between
galaxies/clusters and the large-scale structure surrounding them
\cite[e.g.][]{2001MNRAS.320L...7C, 2004PhRvD..70f3526H,
  2006ApJ...652L..75P, 2007MNRAS.381.1197H, 2007MNRAS.375..184B,
  2011JCAP...05..010B, 2012MNRAS.423..856S,2015MNRAS.448.3391C,
  2015JCAP...08..015B, 2016ApJ...825...49C, 2017MNRAS.468.4502V,
  2017arXiv170608860O, 2017MNRAS.472.1163C, 2017MNRAS.468.4502V,
  2017arXiv170809247B, 2018MNRAS.477.2141O, 2018MNRAS.474.1165P,
  2018MNRAS.481.4753C, 2019MNRAS.485.2492C, 2019A&A...622A..78D}.
This raises the possibility that the orientation of DM haloes is
imprinted in the large-scale structure in the Universe and BCGs
tend to be aligned dynamically toward a particular direction.  Of
course this picture may be over-simplified and should be tested
quantitatively against numerical simulations. This is exactly what we
attempt in the present paper.

Note that a companion paper, \cite{bate2019}, performs a related analysis on
the Horizon-AGN simulation, extending a comprehensive study of
intrinsic alignments by \cite{2017MNRAS.472.1163C}.  While their main
interest lies in the origin of the alignment between elliptical
galaxies and the large-scale structure of the Universe, we investigate
in the present paper the alignments between BCGs and the tidal
fields to understand the origin of the orientations and alignments
between BCGs and their host DM haloes. Therefore, the purposes of our
work and their work are different, although methodology of the
analysis is of course similar.

The structure of this paper is as follows.  Section \ref{sec:method}
first presents a brief summary of our identification scheme of central
galaxies (CGs), simulated counterparts of the observed BCGs, and their
hosting DM haloes from the Horizon-AGN simulation. We also describe
how to estimate the orientation of those objects from an ellipsoidal
fit using a mass tensor, and the tidal field of the large-scale
structure. A representative example for one particular simulated
cluster is shown in Section \ref{sec:example}, and the statistical
analysis over 40 haloes are presented in Section \ref{sec:results}.
Finally, Section \ref{sec:conclusion} is devoted to the summary of the
paper.

The Horizon-AGN simulation adopts the following cosmological
parameters.  The total matter density $\Omega_{m}=0.272$, the baryon
density $\Omega_{b}=0.045$, the dark energy density
$\Omega_{\Lambda}=0.728$, the dimensionless Hubble parameter
$h=0.704$, the amplitude of the power spectrum of density fluctuations
averaged over the sphere of $8h^{-1}$ Mpc radius at present epoch
$\sigma_{8} = 0.81$, the power-law index of the primordial power
spectrum $n_{s}=0.967$.  We also adopt the same cosmological
parameters throughout the paper.

\section{Orientations of dark matter haloes, central galaxies, and
  tidal fields in the Horizon-AGN simulation} \label{sec:method}
The Horizon-AGN simulation \cite[][Papers I and II]{dubois14} follows
the evolution of three components, DM, star, and gas.  DM and stars
are represented by collisionless particles, whereas gas components are
assigned on meshes in the simulation box and solved with the adaptive
mesh refinement. While Papers I and II examined the orientation and
ellipticity of all the three components, this paper focuses on the
relation between DM haloes and CGs, with application to weak lensing
and galaxy surveys in mind. Thus, we do not consider the gas component
in this paper. The Horizon-AGN simulation has a box size of $(100~
h^{-1}$ cMpc$)^3$, where cMpc denotes comoving Mpc.  Since we consider
both small ($\sim$ kpc) and large ($\sim$ Mpc) scales in this paper,
we use both comoving and physical coordinates.  The final mesh size at
the densest region is about $\sim1~h^{-1}$ ckpc.  On the other hand,
dynamics of collisionless dark matter and stellar particles is followed by
the particles-mesh solver \citep{dubois14}.

Since our current study is entirely based on the
  Horizon-AGN simulation, its reliability of baryon physics and the
  extent to which it reproduces the empirical nature of galaxies and
  clusters are crucially important.  It has found to be in good
  agreement, with a number of observed properties including the
  intrinsic alignment of galaxies
  \citep{2015MNRAS.454.2736C,2016MNRAS.461.2702C}, density profile of
  massive galaxies \citep{2017MNRAS.472.2153P, 2019MNRAS.483.4615P},
  cosmic star formation history over the redshift range $1<z<6$
  \citep{2017MNRAS.467.4739K}, morphological diversity of galaxies
  \citep{2016MNRAS.463.3948D}, alignments between
  galactic spin and the nearest filament \cite{2019arXiv190912371W}
  and the ellipticity distribution of X-ray galaxy clusters (Paper
  I). Therefore the Horizon-AGN simulation is supposed to be one of
  the best simulated datasets currently available for our purpose,
  even if not perfect.

\subsection{Identification of
  cluster-sized dark matter haloes and central galaxies
} \label{sec:def}

Following Paper II, we use the ADAPTAHOP \citep{2004MNRAS.352..376A,
  2009A&A...506..647T} to identify DM haloes and galaxies from DM and
stellar particles, respectively, and select cluster-sized haloes with
DM mass of $M_{\rm DM} > 5\times10^{13} ~ M_{\odot}$ at $z\approx0$.
ADAPTAHOP is a subhalo finder that separates multiple subhaloes while
comparing the relative heights of peaks and saddle points of the
smoothed density field.  We select stellar haloes by applying the
ADAPTAHOP to stellar particles.  We define stellar haloes with more
than 50 stellar particles as galaxies.  Since each stellar particles
has the mass of about $2\times10^{6} ~ M_{\odot}$, this criterion
corresponds to the minimum stellar mass of about $10^8 ~ M_{\odot}$ in
our final galaxy catalog.  In total, we have $N_{\rm cl} = 40$ haloes
that are identical to those analysed in Paper II, but we identify
their progenitor haloes at 50 different redshifts so as to trace their
evolution.  The 50 epochs are selected from $z \sim 5$ ($t \sim 1.5$
Gyr) to $z \sim 0$ ($t \sim 13.5$ Gyr) in an equal time interval of
$\Delta t \sim 250$ Myr.  We make the merger trees of all the 40
cluster-sized DM haloes by using TREEMAKER
\citep{2009A&A...506..647T}, which first builds the merger history
tree, and then connects haloes with their progenitors.

Once DM haloes are identified at redshift $0$, we define the CG in each halo as the
most massive galaxy in a halo within 1 pMpc from the most bound
particle of each halo (see Paper II, for more detail).
Thus we define the CG at each epoch $t$ by using the CG in the previous epoch $t-\Delta t$.
Specifically, we define the CG at each epoch $t$ as a galaxy containing the largest number of 
stellar particles of the CG in the adjacent snapshot $t-\Delta t$
and is located within 100 pkpc from the most bound particle of each halo at each epoch $t$.
We expect that the CG selected by the above procedure are similar to observed BCGs.
Finally we define the  "halo centre" by the centre-of-mass of the CG, instead of the centre-of-mass of the DM halo; see equation (\ref{eq:dmtensor}) below.

\subsection{Procedure of ellipsoid fit} \label{sec:ellfit}
Once DM haloes and CGs are identified at each epoch, we fit them to the
triaxial ellipsoid model in three-dimensional space
(Paper I), and measure the major,
intermediate, and minor axis vectors, $\hat{\boldmath{a}}_1$,
$\hat{\boldmath{a}}_2$, and $\hat{\boldmath{a}}_3$, respectively,
unlike in Paper II that fit the data in the projected two dimensional space.

More specifically, we follow the ellipsoid fitting based on the inertia
tensor as described in \cite{2016PASJ...68...97S}.  From all the star
particles belonging to the CG, we first compute its centre-of-mass
position $x_{{\rm CG}, \alpha}^{\rm CM}$ ($\alpha=1, 2, 3$), and
compute the following mass tensor from the star particles located
within a sphere of radius $20$ pkpc from $x_{{\rm CG}, \alpha}^{\rm CM}$:
\begin{equation}
    I_{{\rm CG}, \alpha\beta} (z) \equiv 
    \frac{
     \sum^{N_{\rm star}}_{n=1} m_{\rm star}^{(n)}
     \left[x^{(n)}_{{\rm star}, \alpha} - x_{{\rm CG}, \alpha}^{\rm CM}\right]
     \left[x^{(n)}_{{\rm star}, \beta}   - x_{{\rm CG}, \beta}^{\rm CM}\right]
     }
     {
     \sum^{N_{\rm star}}_{n=1} m_{\rm star}^{(n)}
     },
\label{eq:bcgtensor}
\end{equation}
where $m_{\rm star}^{(n)}$ and $x^{(n)}_{{\rm star}, \alpha}$ are the mass
and the coordinate of the $n$-th stellar particle ($n=1,\cdots, N_{\rm star}$).

The above mass tensor is diagonalized and the directions of the major,
intermediate, and minor axes are computed.  We then select the size of
the ellipsoid $R_{abc}^{\rm star} \equiv \sqrt[3]{a_1a_2a_3}=20$ pkpc,
where $a_1$, $a_2$, and $a_3$ are the half lengths of the major,
intermediate, and minor axes, respectively.  We repeat the above
procedure using the star particles in the ellipsoid around the update
centre-of-mass position $x_{{\rm CG}, \alpha}^{\rm CM}$.
We choose the value of $20$ pkpc as the size of CGs
  for definiteness.  We confirmed that changing the value to $10$ or
  $30$ pkpc does not affect the main conclusion of this paper (see
  also Paper II). 

The whole procedure is iterated until the three eigenvalues of the
mass tensor converge within a fractional error of $10^{-8}$. 
We then {\it redefine} the CG as the set of star particles within the
ellipsoid of $R_{abc}^{\rm star} =20$ pkpc,
and characterize the CG by the parameters including the half lengths of
major axis $a_1$, intermediate axis $a_2$, and minor axis $a_3$ ($a_1
\geq a_2 \geq a_3$), their direction, and the centre of mass $x_{{\rm
    CG}, \alpha}^{\rm CM}$. Therefore the resulting CG is different
from the original set of star particles identified with the ADAPTAHOP
halo finder.

The shape and orientation of the host DM halo at each $z$ are computed
similarly except that we use the mass tensor of DM particles around
the centre-of-mass of the CG:
\begin{equation}
    I_{{\rm DM}, \alpha\beta} (z) \equiv 
    \frac{
     \sum^{N_{\rm DM}}_{n=1} m_{\rm DM}
     \left[x^{(n)}_{{\rm DM}, \alpha} - x_{{\rm CG}, \alpha}^{\rm CM}\right]
     \left[x^{(n)}_{{\rm DM}, \beta}   - x_{{\rm CG}, \beta}^{\rm CM}\right]
     }
     {
     \sum^{N_{\rm DM}}_{n=1} m_{\rm DM}
     }
     ,
\label{eq:dmtensor}
\end{equation}
where $m_{\rm DM}$ and $x^{(n)}_{{\rm DM}, \alpha}$ are the mass
and the coordinate of the $n$-th dark matter particle within the ellipsoid.
In this calculation, we use all the dark matter particles including those in subhaloes.
Once we fix the size of the ellipsoid, $R_{abc}^{\rm DM}
\equiv \sqrt[3]{a_1a_2a_3}$, we can compute the total mass and number
of DM particles within the ellipsoid, $M_{\rm DM}$ and $N_{\rm DM}$.
Unlike in the case of CG, we consider three values of the ellipsoidal bound so that the
corresponding to $M_{\rm DM} = 0.1M_{200}$, $0.5M_{200}$,
and $M_{200}$, where $M_{200}$ is the mass of a sphere whose average
DM density is $200$ times larger than the cosmic critical density at each $z$.

\subsection{The tidal field of the large-scale mass distribution}
\label{sec:tf}

As mentioned in Section \ref{sec:intro}, the orientations of DM haloes are
correlated to their surrounding matter distribution. Let us expand the
gravitational potential of the matter with respect to the centre of
a DM halo, $\bm{x}^{\rm CM}$:
\begin{eqnarray}
 \Phi(\bm{x})
  &=& \Phi(\bm{x}^{\rm CM})
  + \sum_{\alpha=1}^3 (x_\alpha-x_\alpha^{\rm CM})
  \left(\frac{\partial \Phi}{\partial x_\alpha}\right)_{\bm{x}=\bm{x}^{\rm CM}} \nonumber \\
 &&+ \frac{1}{2}\sum_{\alpha,\beta=1}^3
  (x_\alpha-x_\alpha^{\rm CM})(x_\beta-x_\beta^{\rm CM})
  \left(\frac{\partial^2 \Phi}
       {\partial x_\alpha \partial x_\beta}\right)_{\bm{x}=\bm{x}^{\rm CM}} \nonumber \\
  &&+ \cdots .
    \label{eq:Phi-expansion}
\end{eqnarray}
The third term in equation (\ref{eq:Phi-expansion}) describes the tidal
field around the DM halo and is responsible for its ellipsoidal
growth. If we define the tidal field tensor:
\begin{align}
\label{eq:tidal-tensor}
T_{\alpha\beta} =\frac{\partial^2 \Phi}{\partial x_\alpha \partial x_\beta},
\end{align}
its eigen-vectors and eigen-values characterize the direction and
relative growth rate of the ellipsoidal evolution of the object.

We compute the tidal field tensor from the simulation data as follows.
We first divide the simulation box into $100^3$ small grids and assign
the DM density field $\rho(\bm{x})$ at each grid by a cloud-in-cell
interpolation with $\bm x$ being the comoving coordinates of the grid.
Next we define the dimensionless density contrast fields:
\begin{equation}
 \delta(\bm{x}) = \frac{\rho(\bm{x}) - \langle\rho\rangle}{\langle\rho\rangle},
  \label{eq:delta}
\end{equation}
where $\langle\rho\rangle$ is the mean density averaged over the
entire simulation box. Then the tidal tensor $T_{\alpha\beta}(\bm{x})$
at each grid is defined by the second spatial derivative of the
{\it smoothed} density contrast. If we adopt a Gaussian smoothing
over a scale $\sigma$,
the Fourier transform of $T_{\alpha\beta}(\bm{x})$ is easily computed as
\begin{equation}
\tilde{T}_{\alpha\beta} ({\bm k}) = \frac{k_{\alpha} k_{\beta} }{|{\bm{k}}|^2}
\tilde{\delta}({\bm{k}})
\exp\left(\frac{-|{\bm{k}}|^2\sigma^2}{2}\right),
\label{eq:tf}
\end{equation}
where $k_{\alpha}$ and $\tilde{\delta}({\bm k})$ are $\alpha$-th
component of the wave-number vector ${\bm k}$ and the Fourier
transform $\delta({\bm x})$,
respectively.
We use the FFTW package to compute the Fourier transform of the tidal field 
\citep{FFTW98, FFTWgen99, FFTW05}.

Since the spatial extent of cluster-sized haloes is typically $\sim 1
~ h^{-1}{\rm cMpc}$, we choose $\sigma = $3, 5, and 10 $h^{-1}{\rm
  cMpc}$ as the smoothing scale so that the corresponding tidal tensor
traces the large-scale structure surrounding those haloes.  Then, we
compute the inverse Fourier transform of $\tilde{T}_{\alpha\beta}
({\bm k})$ to obtain the tidal tensor $T_{\alpha\beta}(\bm x)$.  We
apply the cloud-in-cell interpolation of the tidal tensors at the
nearby grids to obtain the tidal tensor defined at the centre of the
CG that is assumed to be the centre of the host DM halo as well.

Finally, we diagonalize the tidal field tensor to obtain the normalized
eigenvectors, $\bm \hat{u}_{\alpha}$ ($\alpha=1,2,$ and $3$), and the
corresponding eigenvalues with $\lambda_{1}$, $\lambda_{2}$, and
$\lambda_{3}$ of $\lambda_{1} \geq \lambda_{2} \geq \lambda_{3}$.
 In particular, $\bm \hat{u}_{3}$ corresponds to the
direction of the slowest collapsing or the fastest expanding mode, and
expected to be correlated to the major axis of the object located at
the centre.
Previous studies
\cite[e.g.,][]{2007MNRAS.381...41H,2019ApJ...872...37L} found that
the set of eigenvalues roughly corresponds to the structure defined at
the location as follows;
\begin{description}
\item[(i)] clusters ($\lambda_{1}>0$, $\lambda_{2}>0$, and $\lambda_{3}>0$),
\item[(ii)] filaments ($\lambda_{1}>0$, $\lambda_{2}>0$, and $\lambda_{3}<0$),
\item[(iii)] sheets ($\lambda_{1}>0$, $\lambda_{2}<0$, and $\lambda_{3}<0$),
\item[(iv)] voids ($\lambda_{1}<0$, $\lambda_{2}<0$, and $\lambda_{3}<0$).
\end{description}
We confirmed that for $\sigma = 10$ $h^{-1}$cMpc 11 of our haloes are classified as "clusters",  
and the remaining 29 haloes are as "filaments"
according to the above classification.

\section{An example of ellipsoid fit} \label{sec:example}
\begin{figure*}
\includegraphics[width=\columnwidth]{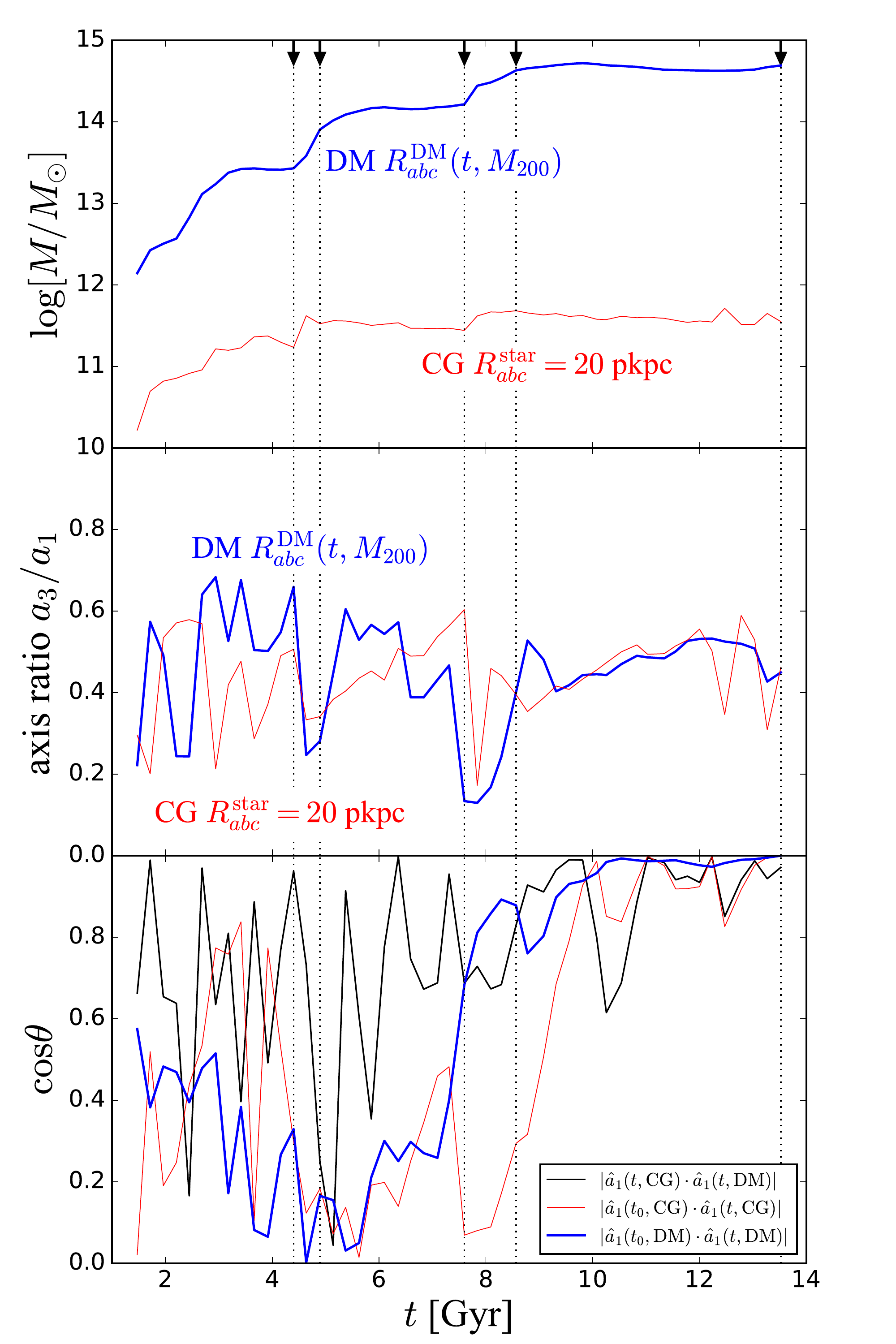}
\caption{ Top: The redshift evolution of the DM mass $M_{200}$ (thick
  blue) and mass of the CG within 20 pkpc (thin red) for an example of
  one halo shown in Figure \ref{fig:z}.  Vertical dotted lines
  correspond to five epochs shown in Figure \ref{fig:z}.  Middle: The
  redshift evolution of major-to-minor axis ratios $a_3/a_1$ of fitted
  ellipsoids both for the DM and for the CG.  Bottom: Alignment angles
  between orientations of the CG and the DM halo at each
  epoch (black), orientations of the DM at the present epoch and in
  the past (thick blue), and orientations of the CG at the present
  epoch and in the past (thin red). }
    \label{fig:t_ar_no2}
\end{figure*}

\begin{figure*}
\includegraphics[width=2\columnwidth]{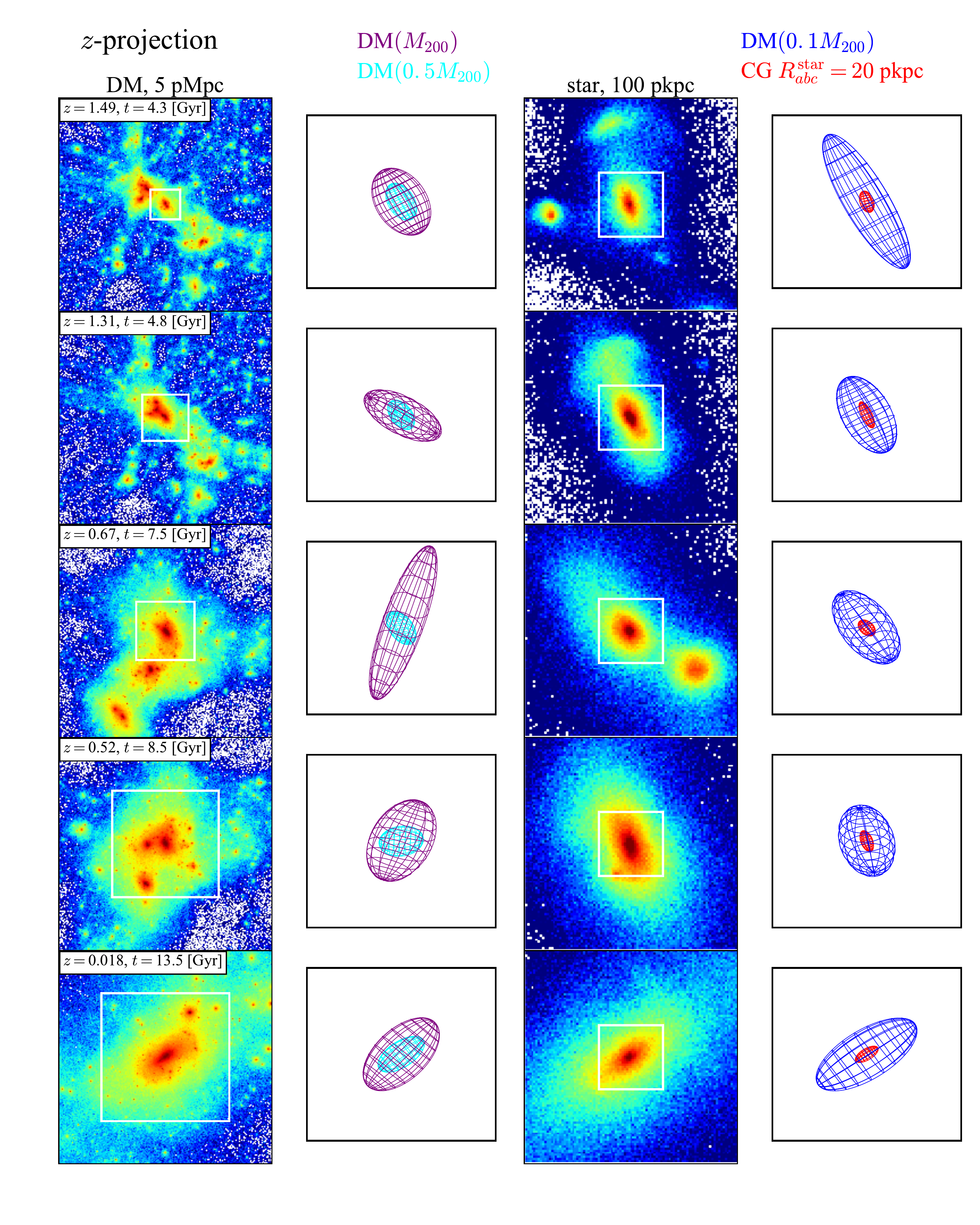}
\caption{From left to right, we show images projected along the
  $z$-direction of the Horizon-AGN simulation box of DM
  particles within a (5 pMpc)$^3$ cube,
  fitted ellipsoids of DM for the enclosed mass
  of $M_{200}$ (purple) and $0.5M_{200}$ (cyan), 
  stellar particles within a (100 pkpc)$^3$ cube,  
  and the ellipsoids of the DM for $0.1M_{200}$ (blue) and CG (red), respectively.  
  From top to bottom, the images correspond to those at $z=1.49{~}(t=4.3 ~ {\rm
      Gyr})$, $z=1.31 {~}(t=4.8 ~{\rm Gyr})$, $z=0.67 {~}(t=7.5
  ~ {\rm Gyr})$, $z=0.52{~}(t=8.5~{\rm Gyr})$, and $z=0.018
  {~}(t=13.5~{\rm Gyr})$, respectively.  These five epochs are also
  indicated by vertical dotted lines in Figure \ref{fig:t_ar_no2}.  }
    \label{fig:z}
\end{figure*}

In this section, we select the most massive single-core-dominated halo
from the 40 haloes in our sample, which is the same as plotted in
Figure 3 of Paper I.  Figure \ref{fig:t_ar_no2} shows the evolution of
the mass of the DM halo and CG (top), the ratios of their major and
minor axes $a_3/a_1$ (middle), and the angles between their major axes
(bottom).  The masses of the DM halo and CG are
plotted in blue and red, respectively, in the top panel. The axis
ratios, $a_3/a_1$, are computed for the ellipsoids enclosing those
masses and plotted in the same colour, respectively.  The bottom panel
plots absolute values of the three direction cosines of the different
major axes.  The black line is computed from ${\hat{a}_1}$ of the CG
and the DM halo at the same epoch $t$. The red and blue lines are
computed from ${\hat{a}_1}$ defined at $t$ and the present epoch $t_0$
for the CG and the DM halo, respectively.

We choose five redshifts (indicated by the vertical dotted lines) to
investigate the snapshots in more detail; before and after two major
merger events ($z=1.49$, $1.31$, $067$, and $0.52$) and at present
($z\approx0$).  The signature of the mergers is clearly seen in the
top panel of Figure \ref{fig:t_ar_no2}, where the DM halo mass
significantly increases.  
The first and second columns in Figure \ref{fig:z} show
the surface density of dark matter component and the corresponding
ellipsoids projected along the $z$-axis, respectively. 
At each redshift, we
extract a cube of $(5 ~{\rm pMpc})^3$ around the centre of the CG of that halo.  
The white squares in the first column indicate the box square in the second
column.  Similarly, we extract a cube of $(100 ~{\rm pkpc})^3$
around the centre of the CG, and plot the surface density of stellar
component and the corresponding ellipsoids in the third and fourth
columns, respectively.

Figure \ref{fig:t_ar_no2} indicates that masses, axis ratios, and
orientations of those objects did not change much after the last major
merger around 8 Gyr. Before the epoch, the axis ratios and the
orientations change significantly, presumably due to repeated
mergers or mass accretion events during the growth of the halo.  In
particular, the shape of the DM halo became very elongated at the two
major merger events, leading to rapid changes of $a_1$ during the
mergers.  This also leads to the enhancement of the angular momentum
amplitude during the merger episode \citep{2004MNRAS.348..921P}.  While there
are large variations between the orientations of the CG and the host
DM halo, they are relatively well aligned at each epoch (black line in
the bottom panel), and evolve coherently toward their current
direction (blue and red lines).

The above features are visually illustrated in Figure \ref{fig:z}.
The major merger between $z=1.49$ and $1.31$ proceeded through the
mass accretion along the upper-left to lower-right filamentary
structure.  Thus the major axes of the DM haloes and CG follow the
direction of the filament and do not change much, even though their
ellipticities, in particular at the outer boundary, significantly
change during the merger event. A similar trend is seen at the next
major event between $z=0.67$ and $0.52$.

After $z=0.52$ ($t=8.5$ Gyr), the DM halo did not experience any violent
merger (see the top panel of Figure \ref{fig:t_ar_no2}), and the
axis ratio and direction of the major axis of the outer boundary of
the DM halo (corresponding to $M_{200}$) are fairly constant until the
present epoch. The orientations of the inner DM haloes defined at
$0.1M_{200}$ and $0.5M_{200}$ and the CG gradually became aligned
toward that of the outer DM halo.

The evolution history of this specific halo presented in Figures
\ref{fig:t_ar_no2} and \ref{fig:z} is summarized as follows. The DM
halo grows through sequences of repeated mergers and mass
accretions following the surrounding large-scale structure, in
particular along the nearby filamentary structures. The shape and
orientation of the DM halo are significantly affected by those events,
whereas they did not evolve much after the last major merger around 8
Gyr. The inner part of the DM halo and CG evolves rather coherently so that
their major axes become aligned better toward that of the outer DM
halo, which is basically fixed just after the last major event.

It is not clear, however, to what extent the above simple picture is applicable
to other DM haloes and CGs as well in general. Therefore we analyse the
orientations of all the 40 haloes and study the statistical
evolution behavior in the next section.

\section{Statistical correlation among orientations of
  DM haloes, CGs, and surrounding tidal field} \label{sec:results}

In order to examine the validity of a simple picture emerging from the
evolution of the particular halo presented in the previous section,
we consider three different aspects of the statistical correlation
over 40 simulated haloes; (i) instantaneous correlation of
orientations between CGs and DM haloes, (ii) evolution of the
orientation of CGs and DM haloes towards their present values, and
(iii) statistical correlation and evolution of their orientation with
respect to the surrounding tidal field.  As we will show below,
those results indicate that the orientations of DM haloes at the present epoch are
basically imprinted in the initial conditions of the large-scale structure, 
while the orientations of CGs drastically evolve with time due to mergers and mass accretions.

\subsection{Instantaneous correlation of
orientations between the CGs and DM haloes}
\label{sec:t_cos_bcg}

We first examine to what extent the orientations of CGs are aligned
to that of the host DM haloes instantaneously. For that purpose, we
compute the direction cosines between the unit vectors along the
major axes of CGs and DM haloes at the same epoch, and then  
average them  over the entire 40 haloes:
\begin{equation}
  \langle\cos\theta\rangle(t; {\rm CG}-{\rm DM})
  \equiv \frac{1}{N_{\rm cl}}\sum_{i=1}^{N_{\rm cl}}
         \left|{\bm \hat{a}_1}^{(i)}(t, {\rm CG})
         \cdot {\bm \hat{a}_1}^{(i)}(t, {\rm DM})\right|.
 \label{eq:mean-cos-t-CG-DM}
\end{equation}

\begin{figure*}
	\includegraphics[width=\columnwidth]{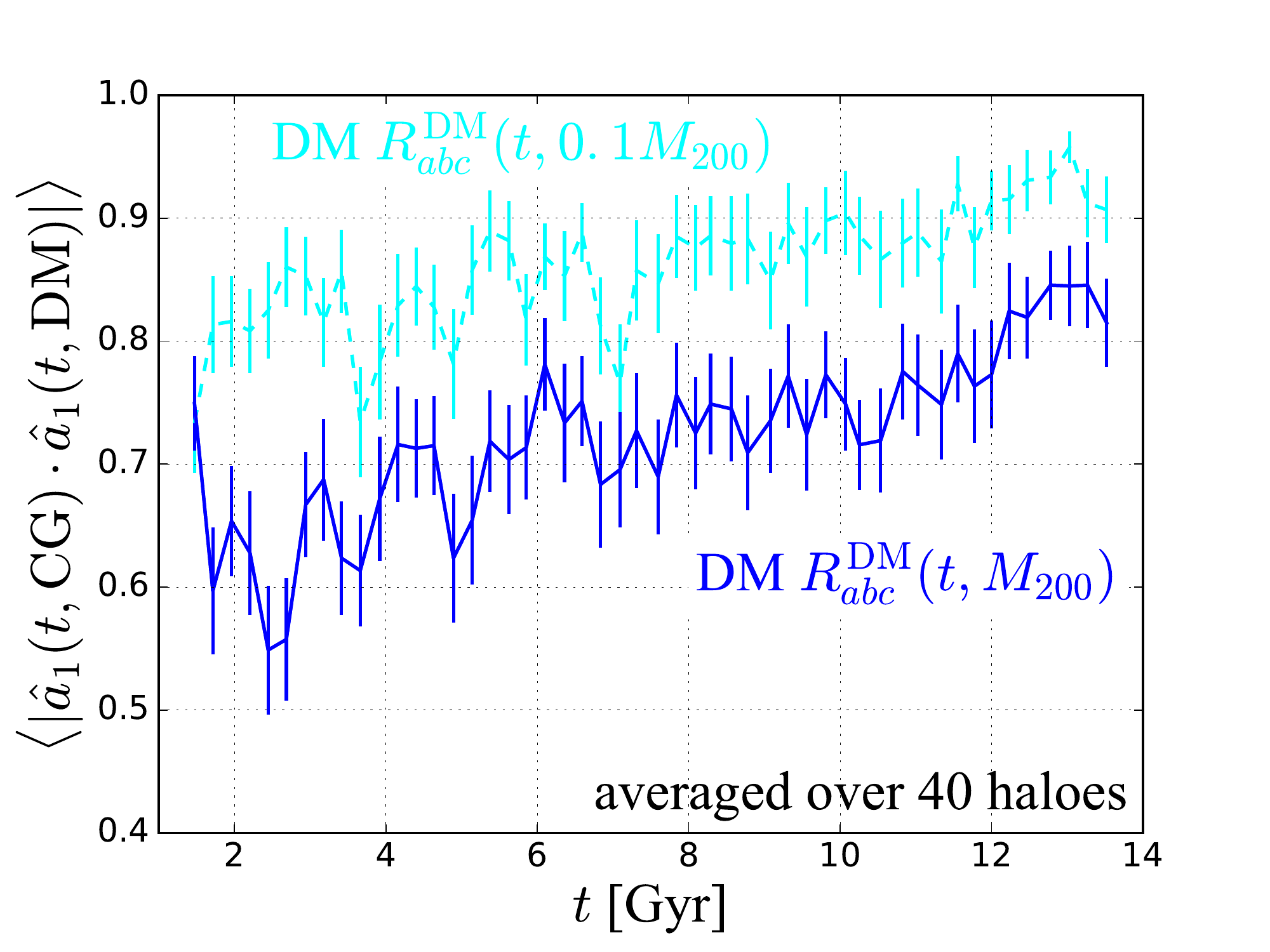}
\caption{
  Correlation between orientations of CGs and DM haloes
  evaluated at the same epoch.  Dashed cyan and solid blue lines
  indicate the direction cosine between CGs and DM haloes for enclosed
  masses of $0.1M_{200}$ and $M_{200}$, respectively, averaged over 40
  haloes. 
  The error bars correspond to the determination accuracy of the mean
  values defined as the standard deviation divided by the square root
  of the number of haloes, $N_{\rm cl}=40$.}
     \label{fig:mean-cos-t-CG-DM}
\end{figure*}

\begin{figure*}
	\includegraphics[width=1.8\columnwidth]{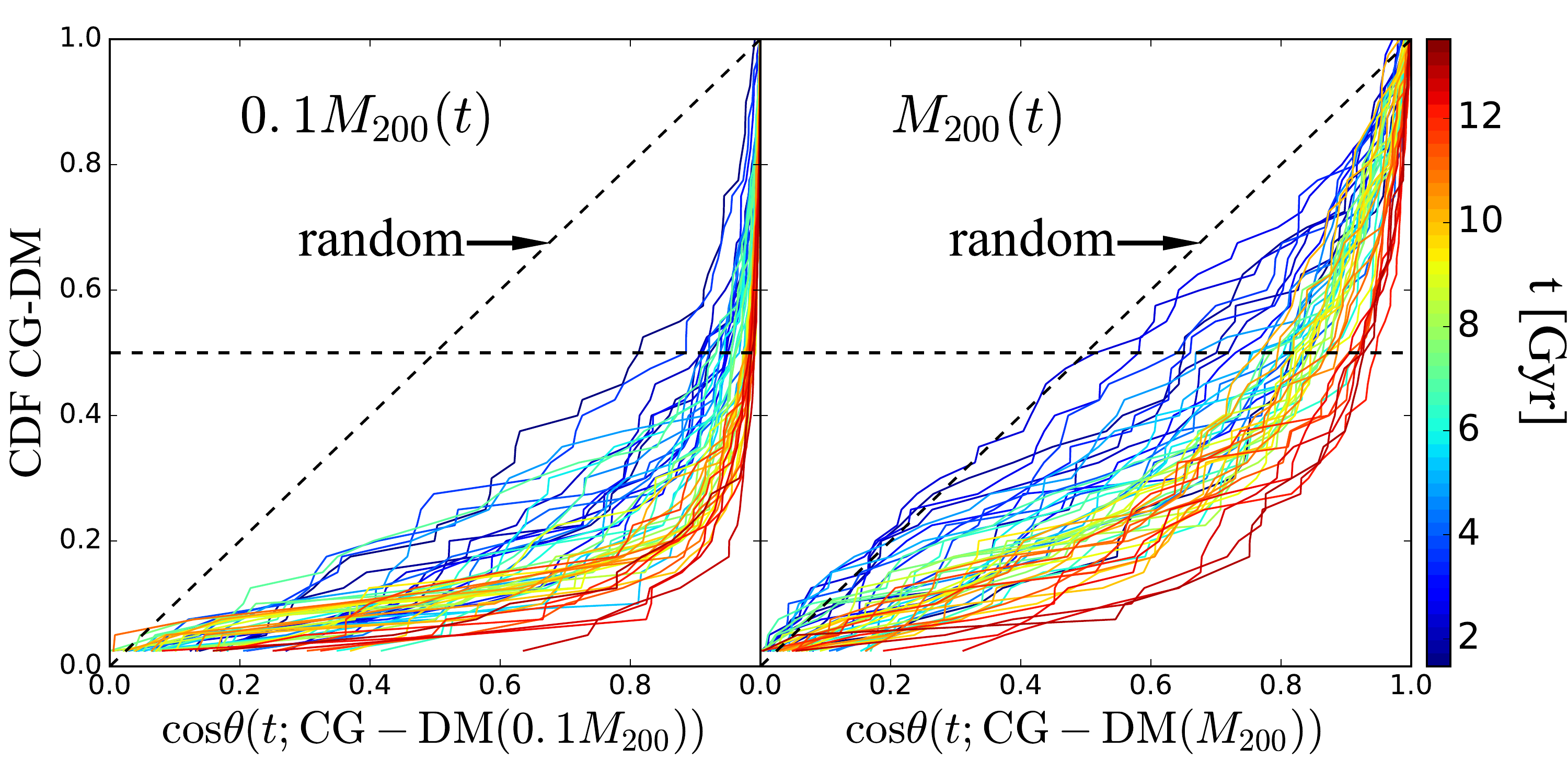}
\caption{ 
  Cumulative probability distributions of alignment angles
  between orientations of DM haloes and CGs at each epoch $t$.  Left and right
  panels show results for DM haloes with the enclosed mass
  $0.1M_{200}$ and $M_{200}$, respectively.   Colour
  scale corresponds to the cosmic time, bluer lines are earlier and
  redder lines are later.}
    \label{fig:cdf_dm_bcg}
\end{figure*}

Figure \ref{fig:mean-cos-t-CG-DM} plots equation
(\ref{eq:mean-cos-t-CG-DM}) for CGs against their host DM haloes
defined at the mass scale of $M_{200}$ (blue-solid line) and
$0.1M_{200}$ (cyan-dotted line). Since equation
(\ref{eq:mean-cos-t-CG-DM}) should reduce to $0.5$ (or
$\cos^{-1}(0.5)=60^\circ$) if the two major axes are uncorrelated and
randomly oriented, Figure \ref{fig:mean-cos-t-CG-DM} indicates that
the major axes of CGs are always positively aligned to those of their host DM haloes.
In order to see the evolution of the above alignment more clearly, we
plot their cumulative probability density functions in Figure
\ref{fig:cdf_dm_bcg} at 50 epochs.
The alignment between CGs and their host DM haloes
becomes more tightly aligned toward the present epoch.

As expected, CGs are correlated more strongly with the inner part of
the DM haloes at any epoch, with a mean relative angle less than
$\cos^{-1}(0.8)\approx 40^\circ$.  This result is qualitatively
consistent with the observational claim by \cite{2017NatAs...1E.157W}
that orientations of BCGs and their host DM haloes are aligned even at $z>1.3$ ($t<5$ Gyr).
It is not easy, however, to compare our results with \cite{2017NatAs...1E.157W}
quantitatively, partly because 
cluster masses of the \cite{2017NatAs...1E.157W} sample are $M_{\rm vir} \sim 10^{15}~M_{\odot}$,
whereas masses of our sample are $M_{200} \sim 10^{14}~M_{\odot}$.
We also find that the correlation with DM haloes increases gradually on average toward the present
epoch. 
The average alignment angles between CGs and the outer boundary of DM haloes
at $M_{200}$ are $\approx \cos^{-1}(0.70) = 45^\circ$ before $t=8$Gyr
and $\approx \cos^{-1}(0.82) = 35^\circ$ at present (see also Paper II), respectively.

\begin{figure*}
	\includegraphics[width=\columnwidth]{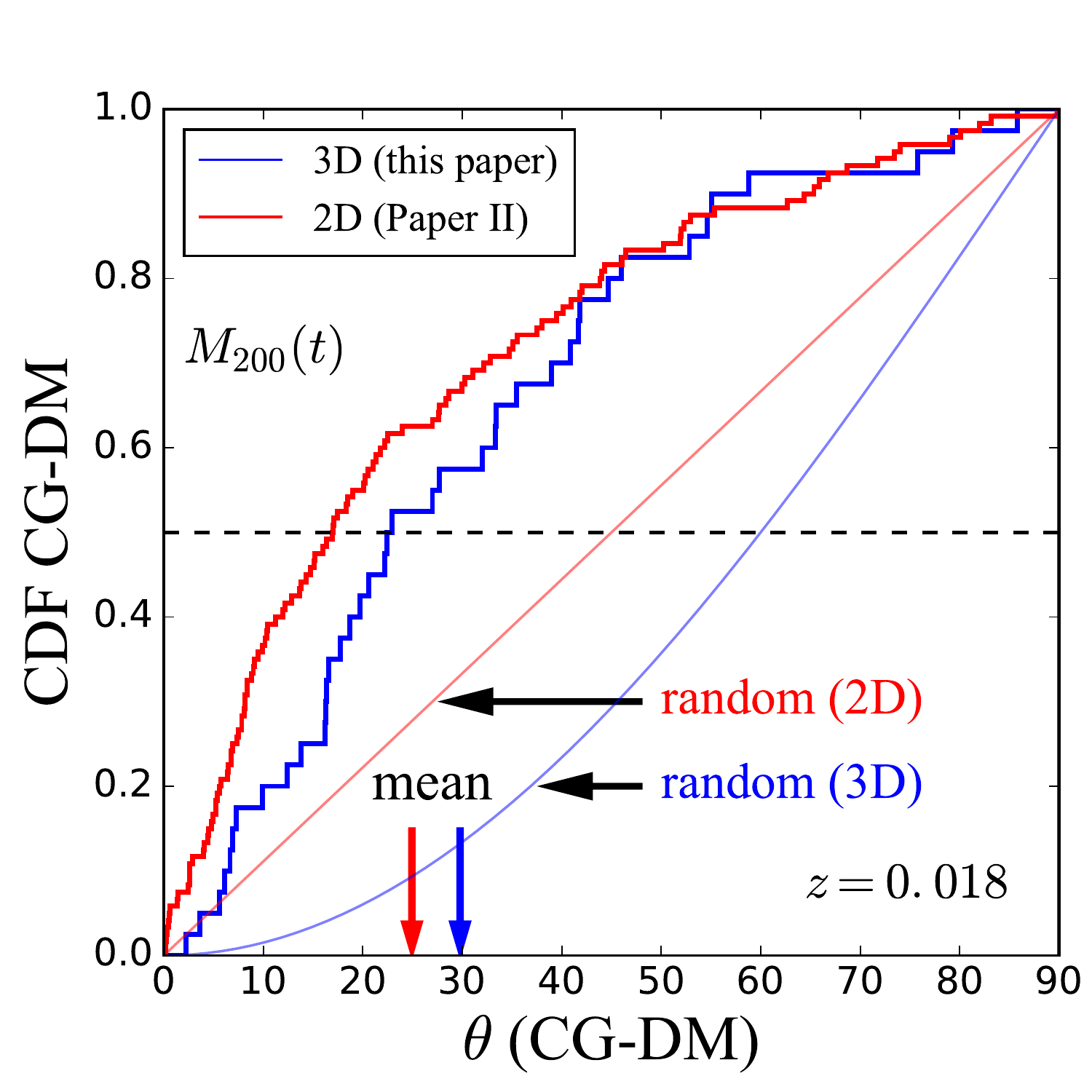}
\caption{ Cumulative probability distributions of alignment angles $\theta$
  between orientations of DM haloes and CGs at present epoch $z=0.018$. 
  Blue and red thick lines correspond to the alignment angles of three dimensional (3D)
  fit (this paper) and two dimensional (2D) fit (Paper II), respectively.
  The angles $\theta$ of 3D fit is the same as the right panel in Figure \ref{fig:cdf_dm_bcg}.
  The angles $\theta$ of 2D fit is the same of the bottom right panel in Figure 6 of Paper II.
  Red and blue thin lines correspond to the cumulative probability distribution of the random distributions.
  Mean values of $\theta$ for both the 3D and 2D are shown with arrows.
   }
    \label{fig:hist_2d_3d}
\end{figure*}

Since the angles are observationally measurable only in the projected
two dimensional plane, Figure \ref{fig:hist_2d_3d} compares the
cumulative distribution of the angles defined in three dimensional
space (see Figure \ref{fig:mean-cos-t-CG-DM}) with those similarly
defined after projected along either $x, y, $ or $z$ direction in the
simulation coordinates at $z\approx0$.  This plot helps understanding
the connection between the three dimensional angles studied in this
paper and observable two dimensional angles.

\subsection{Evolution of
  orientations of CGs and DM haloes towards the present time}
\label{sec:t_cos_avg}
\begin{figure*}
\includegraphics[width=\columnwidth]{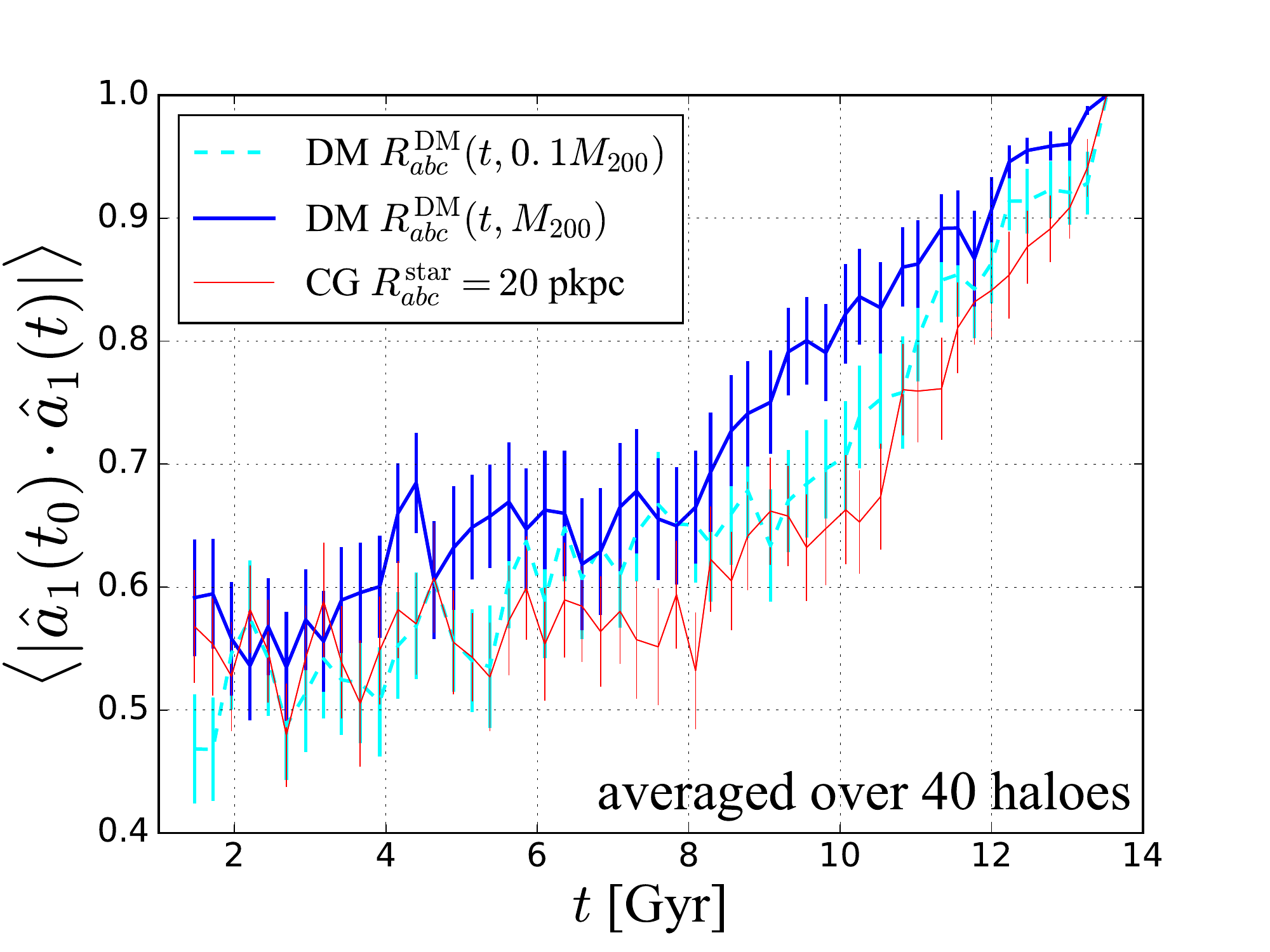}
\caption{Correlation between orientations of objects at $t$ and the
  present epoch $t_0$ for the three components; Red-thin line is
  for CGs, and cyan-dashed and blue-solid lines are for DM haloes with
  enclosed masses of $0.1M_{200}$ and $M_{200}$, respectively. 
  The quoted error-bars represent the root mean square value 
  divided by $\sqrt{N_{\rm cl}}$.
  }
    \label{fig:mean-cos-t-t0}
\end{figure*}

We consider next how the orientations of CGs and DM haloes become
aligned towards their present values. Figure \ref{fig:mean-cos-t-t0} plots
\begin{equation}
  \langle\cos\theta\rangle(t,t_0; {\rm X})
  \equiv \frac{1}{N_{\rm cl}}\sum_{i=1}^{N_{\rm cl}}
         \left|{\bm \hat{a}_1}^{(i)}(t, {\rm X})
         \cdot {\bm \hat{a}_1}^{(i)}(t_0, {\rm X} )\right|
 \label{eq:mean-cos-t-t0}
\end{equation}
for the three components, X=CG (red-thin solid) and DM haloes of
$0.1M_{200}$ (cyan-dashed) and $M_{200}$ (blue-solid).

Orientations of the major axes of those objects at early epochs
($t\leq4$ Gyr) are quite different from the ones at the present time; the
average alignment angles $\theta(t, t_0)$ are somewhere between
$50^\circ$ and $60^\circ$, corresponding to $\cos^{-1}(0.6)$ and
$\cos^{-1}(0.5)$.  This result confirms the scenario presented in
Section \ref{sec:example}: orientations of both DM haloes and CGs
change drastically with time.  The correlation of each component
increases gradually and steadily toward the present epoch, in
particular, at $t>8$ Gyr.

Since Figure \ref{fig:mean-cos-t-t0} may suggest a
  possible break of the correlation curves around $t=8$ Gyr, we
  examined both the occurrence rate of the last major merger events
  and the cluster mass growth history for the 40 haloes
  individually. However, they seem to be fairly continuous around
  $t=8$ Gyr, and therefore we do not think that this epoch has any
  particular physical meaning.  On the other hand, it corresponds
  approximately to the median epoch when the mass of each cluster
  exceeds the half of its current value.
  This may explain why orientations of both DM haloes and CGs
  remain close to their present ones at $t>8$ Gyr.

Figure \ref{fig:mean-cos-t-t0} also appears to indicate that the
orientations of the outer DM haloes first become aligned closer to its
present value, followed by that of the inner DM haloes, and then by
that of CGs.  This result suggests that the alignment
proceeds from larger to smaller scales.  Therefore those orientations
and their mutual alignment may be determined by the surrounding
larger-scale structure.

\begin{figure*}
\includegraphics[width=\columnwidth]{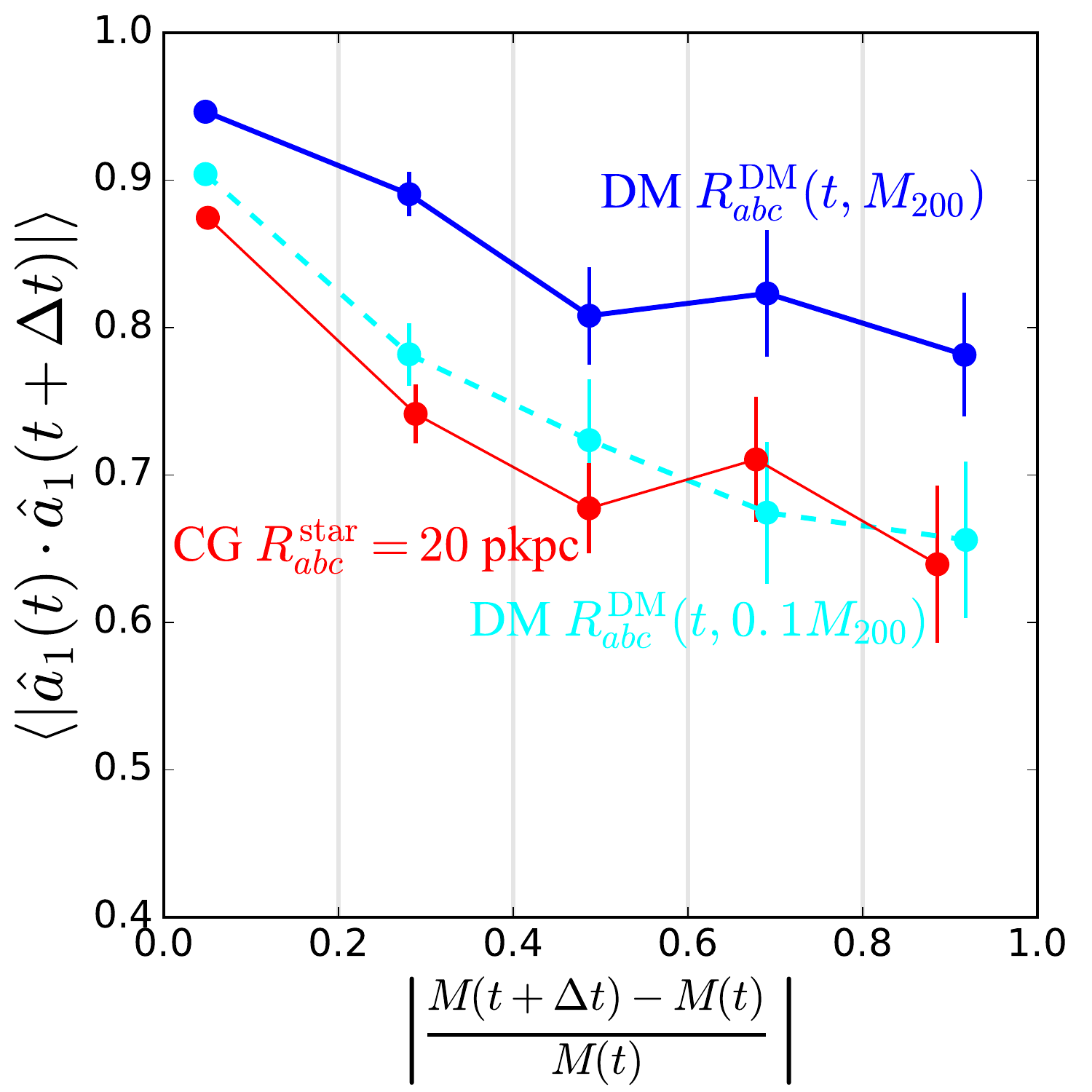}
\caption{
Correlations between changes of orientations and  those of masses during the time interval $\Delta t = 250$ Myr.
We take the absolute values of the fractional mass differences because large negative values correspond to 
flyby galaxies that are below the detection threshold of the galaxy finder and detach after their passage, and therefore negative values are similar to mergers with large positive values.
For each bin of the fractional mass difference, we show the average and error of direction cosines of major axes of neighboring epochs.
To compute the averages and the errors, we use all the 40 haloes and 49 snapshot pairs for each halo.
The quoted error-bars represent the root mean square value divided by the square root of the number of corresponding objects in each bin.
Red-thin line is for CGs, and cyan-dashed and  blue-solid lines are for DM haloes with enclosed masses of $0.1M_{200}$ and $M_{200}$, respectively.
}
    \label{fig:cos_dM}
\end{figure*}

Figure \ref{fig:mean-cos-t-t0} implies that the change of orientations
between DM haloes and CGs is driven by strong dynamical
interactions through successive mergers and mass accretion episodes.  To check
this point more explicitly, in Figure \ref{fig:cos_dM} we show the
correlation between fractional mass changes and changes of
orientations at neighboring snapshots with a time interval of $\Delta
t =250$ Myr.  Figure \ref{fig:cos_dM} indicates that changes of
orientations are large when fractional mass differences are large,
which correspond to mergers and large mass accretions, both for DM
haloes and CGs.
This suggests that the spin swings of both DM haloes and
  galaxies are mainly driven by their mergers and mass accretions,
  while they are also affected by the later re-distribution of the
  angular momentum vector inside them.  This picture is qualitatively
  consistent with the result of \citet{2014MNRAS.445L..46W}. 

\subsection{Orientations of DM haloes and CGs with respect to
the surrounding large-scale structure} 
 \label{sec:t_cos_cluster_tf}
The results presented in the previous subsections imply that the large
scale environment is responsible for the
orientations and the alignments of CGs and their host DM
haloes. Thus we choose the orientations of the eigenvectors of the tidal
field as a proxy of the directions embedded in the large scale
structure, which may keep the memory of the initial conditions.

\begin{figure*}
\includegraphics[width=1.8\columnwidth]{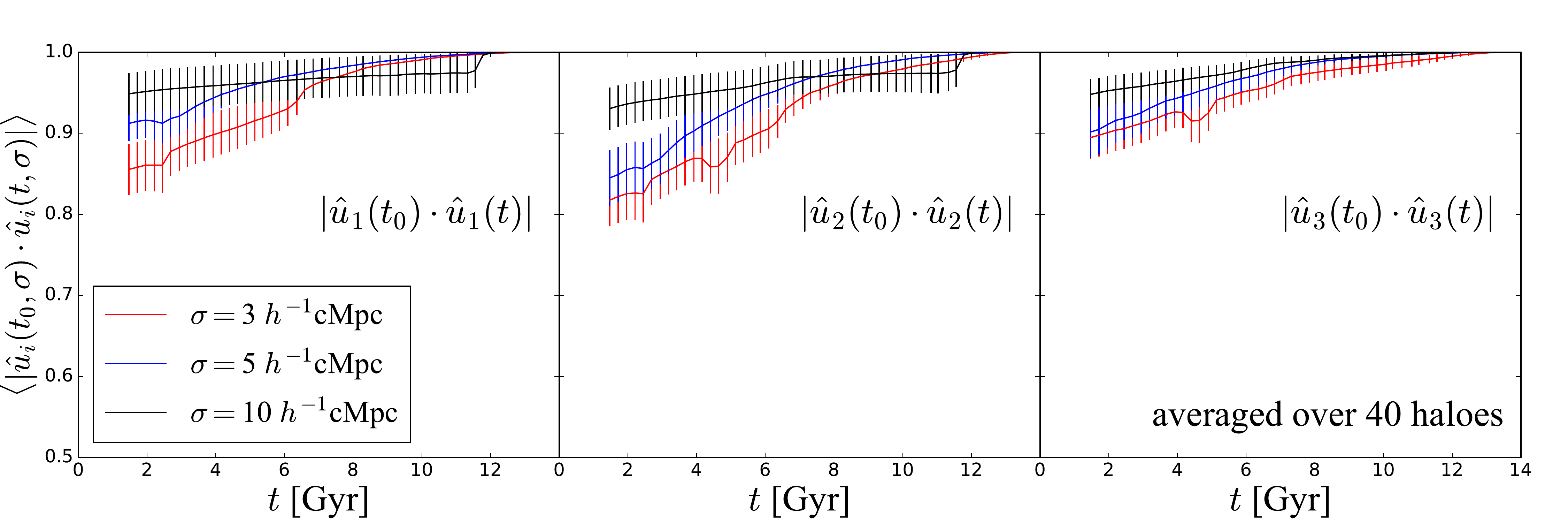}
\caption{ Correlation between the eigenvectors of the tidal field at
  $t$ and the present epoch $t_0$. They are computed from a density
  field Gaussian-smoothed over $\sigma = 3h^{-1}$ (red), $5h^{-1}$
  (blue), and $10h^{-1}$ (black) cMpc; see Section \ref{sec:tf} for
  further details.  The eigenvectors are labelled as ${\bm
    \hat{u}_{1}}$, ${\bm \hat{u}_{2}}$, and ${\bm \hat{u}_{3}}$
  corresponding to the largest, medium, and smallest eigenvalues.
  Their correlations $\left|{\bm \hat{u}_{\alpha}}(t_0) \cdot {\bm
    \hat{u}_{\alpha}}(t)\right|$ averaged over the 40 halo locations
  are plotted for $\alpha=1$, 2 and 3 in the left, centre and right
  panels, respectively.
  The quoted error-bars represent the root mean square value 
  divided by $\sqrt{N_{\rm cl}}$.
  The sudden change at $\sim12$ Gyr in the left and middle panels 
  is due to an outlier cluster whose eigenvectors suddenly change at that epoch.
  }
    \label{fig:t_cos_avg_tf_tf}
\end{figure*}

Figure \ref{fig:t_cos_avg_tf_tf} plots the correlation of the three
eigenvectors ${\bm \hat{u}}_\alpha$ computed at each epoch ($t$) and the present
epoch ($t_0$).  We apply three different smoothing lengths, $\sigma =
3h^{-1}$, $5h^{-1}$, and $10h^{-1}$ cMpc, and compute the eigenvectors
at the location of CGs according to the procedure described in
Section \ref{sec:tf}.  As is clear from Figure
\ref{fig:t_cos_avg_tf_tf}, those eigenvectors do not change so much over
the cosmic time.

In particular, directions of the tidal field eigenvectors with
$\sigma=10~h^{-1}$cMpc are fairly constant over $\sim10$ Gyr.
Since $10$  $h^{-1}$cMpc is sufficiently larger than the size of the typical
cluster-sized haloes and less than the typical separation ($\sim30$ $h^{-1}$cMpc) of the nearest
cluster-sized halo, we choose $10h^{-1}$ cMpc as the smoothing length
in the following analysis, and adopt ${\bm
  \hat{u}}_\alpha(t_0;\sigma=10 ~ h^{-1}{\rm cMpc} )$ defined at the
CG's location as a set of proxies for the preferential directions
imprinted in the large-scale structure surrounding those haloes.

\begin{figure*}
\includegraphics[width=1.8\columnwidth]{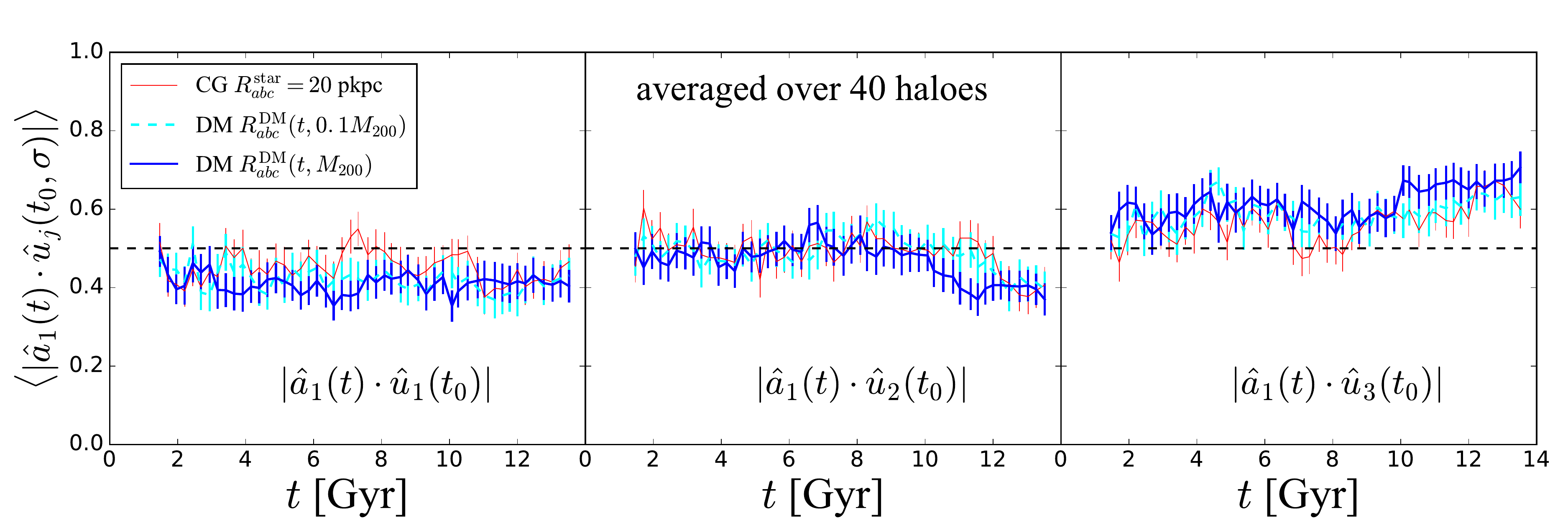}
\caption{ Mean values of alignment angles between orientations of
  haloes at each epoch $t$ and eigenvectors of the tidal field at
  the present epoch $t_0$.  Left, middle, and right panels show the
  alignments of halo orientations with respect to eigenvectors
  ${\bm \hat{u}_{1}}$, ${\bm \hat{u}_{2}}$, and ${\bm \hat{u}_{3}}$, respectively.
  Dashed cyan, thick blue, and thin red lines indicate median
  alignment angles of DM haloes for enclosed masses of
  $0.1M_{200}$, $M_{200}$, and those of CGs, respectively.  
  The quoted error-bars represent the root mean square value 
  divided by $\sqrt{N_{\rm cl}}$.
  The smoothing scale of
  the tidal field is set to $\sigma=10~h^{-1}$cMpc.  }
    \label{fig:t_cos_avg_dm_tf}
\end{figure*}

In order to see the relation of the orientations of objects and the
surrounding environment, we compute the correlations of the major axis
direction of ${\bm \hat{a}_1}(t; {\rm X})$, where $X={\rm CG}$, inner
DM halo, and outer DM halo, against the eigenvectors of the tidal field
${\bm \hat{u}}_\alpha(t_0)$ averaged over the 40 halo locations.
Figure \ref{fig:t_cos_avg_dm_tf} plots $\langle\left|{\bm
  \hat{a}_1}(t) \cdot {\bm \hat{u}}_\alpha(t_0) \right|\rangle$
as a function of $t$ for $\alpha=1$, 2, and 3 in the left, centre, and right
panels, respectively. Each panel has three curves corresponding to the
three objects; CG (red), inner DM halo (cyan) and outer DM halo (blue).
The major axes of the three objects exhibit positive and negative
correlations with ${\bm \hat{u}}_3(t_0)$ ($\sim0.6$) and ${\bm
  \hat{u}}_1(t_0)$ ($\sim0.4$), respectively, relative to the random distribution.
The intermediate axis of the tidal field, on the contrary, is almost
uncorrelated ($\sim0.5$) with the major axis of the objects, although they tend to
become weakly negative correlated gradually toward the present epoch ($\sim0.4$).  

\begin{figure*}
	\includegraphics[width=1.8\columnwidth]{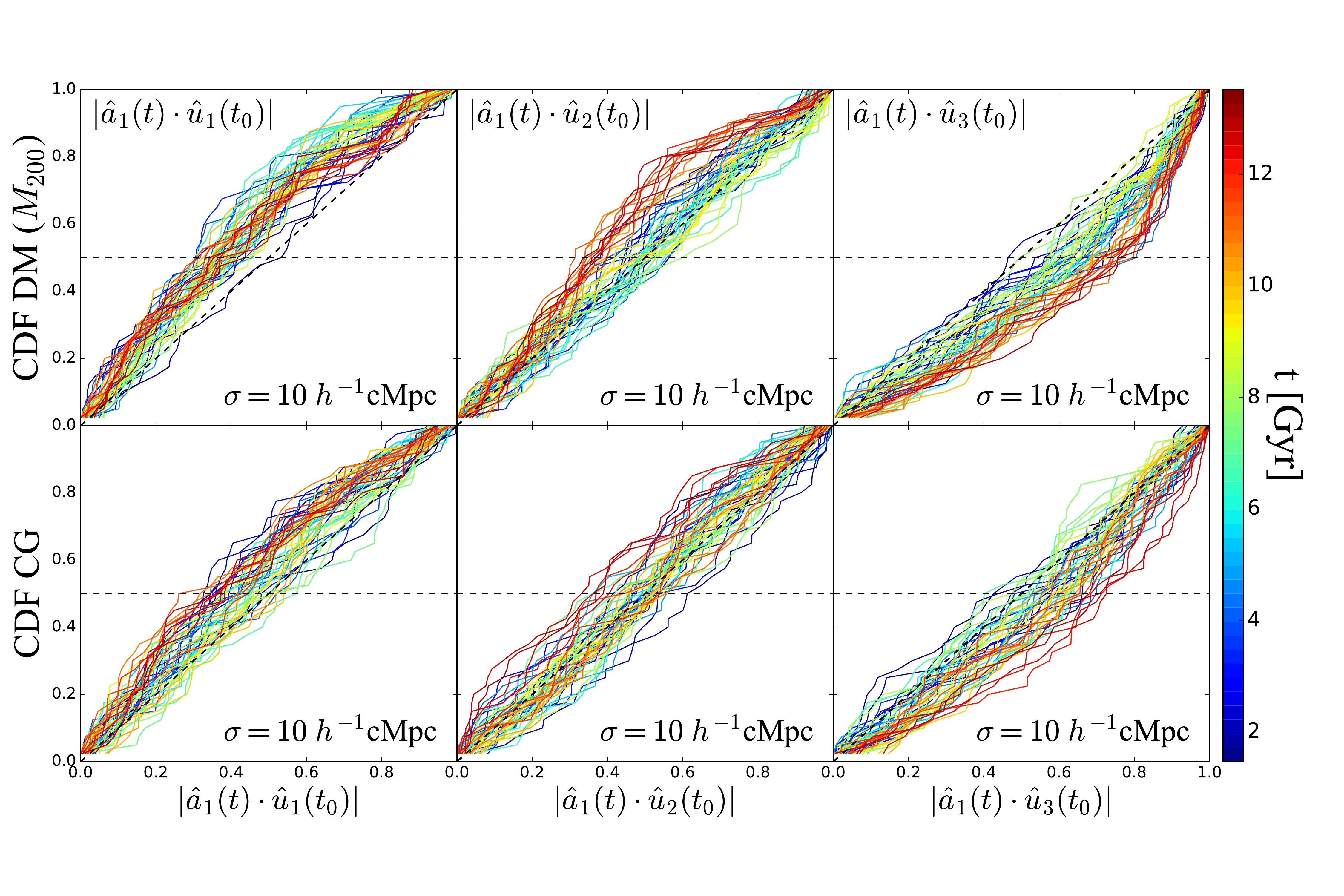}
\caption{ Cumulative probability distributions of alignment angles
  between orientations of haloes at each epoch $t$ and eigenvectors of
  the tidal field at the present epoch $t_0$.  Top and Bottom panels
  show results for dark matter haloes with the enclosed mass $M_{200}$
  and for CGs, respectively.  Left, middle, and right panels show the
  position angles of ${\bm \hat{u}_{1}}$, ${\bm \hat{u}_{2}}$, and
  ${\bm \hat{u}_{3}}$ relative to ${\rm a_1} (t, {\rm DM})$,
  respectively.  Colour scale corresponds to the cosmic time, bluer
  lines are earlier and redder lines are later.  The smoothing scale
  of the tidal field of $\sigma=10$ $h^{-1}$cMpc is adopted.  }
    \label{fig:cdf}
\end{figure*}

In order to see the evolution of the above alignment more clearly, we plot the cumulative probability density functions in Figure \ref{fig:cdf}. 
The upper and lower panels show those for DM haloes and CGs against $\bm{\hat{u}}_{1}$ (left), $\bm{\hat{u}}_{2}$ (centre), and $\bm{\hat{u}}_{3}$ (right). 
Each curve represents the cumulative probability density function at $t$ according to the colour-bar shown to the right. 
The diagonal dotted line indicates the completely random distribution.
Positive and negative correlations correspond to the convex and concave curves in Figure \ref{fig:cdf}, respectively.

As we have seen in Figure \ref{fig:t_cos_avg_dm_tf}, the major axes of
DM haloes evolve preferentially toward the direction of
$\bm{\hat{u}}_{3}(t_0)$.  The major axes of DM haloes tend to be away
from $\bm{\hat{u}}_{1}(t_0)$ in a time-independent manner.  They are
fairly uncorrelated with $\bm{\hat{u}}_{2}(t_0)$ at the early epochs,
but develop weak correlation toward the present epoch.  The
correlation of CGs against the tidal field are weaker than that of DM
haloes, but exhibits qualitatively a similar trend.  This is
consistent with the fact that 11 and 29 out of our 40 clusters
correspond to ``clusters'' and ``filaments'', respectively, according
to the definition in Section \ref{sec:tf}
\cite[e.g.][]{2007MNRAS.381...41H}.

\cite{bate2019} have studied in particular the evolution of alignments
of massive elliptical galaxies relative to the tidal field.  They
find that the alignments are tighter for ${\bm \hat{u}_{1}}$ and ${\bm \hat{u}_{3}}$
than for ${\bm \hat{u}_{2}}$, and also that the
alignments increase from $z = 3$ to $0$.  These two findings
are consistent with our results.

\section{ Summary } \label{sec:conclusion}
\begin{figure*}
	\includegraphics[width=1.5\columnwidth]{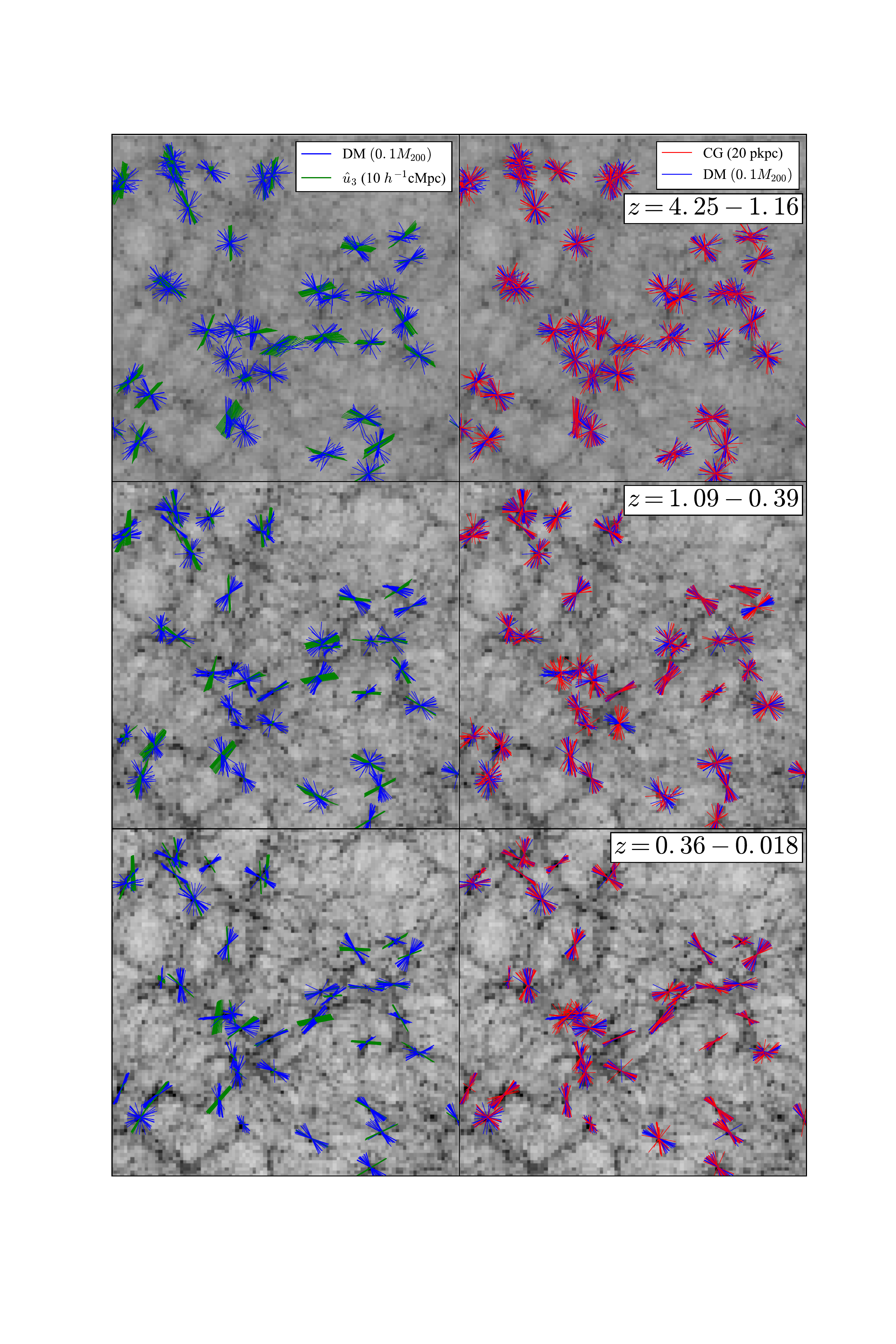}
\caption{ 
  Projected mass density fields of DM component and the orientations of
  CGs (red), DM haloes for the enclosed mass
  of $0.1M_{200}$ (blue), and the tidal field eigenvectors ${\bm \hat{u}_{3}}$
  (green) for early (top panel, $z=4.25$-$1.16$, $t =$ 1.5-5.4 Gyr), middle (middle
  panel, $z=1.09$-$0.39$, $t =$ 5.6-9.6 Gyr), and late (bottom panel, 
  $z=0.36$-$0.018$, $t =$ 9.8-13.5 Gyr) epoch.
   In each panel, all the eigenvectors in the redshift range are shown.
  The size of each panel corresponds to the simulation
  box size, $100$ $h^{-1}$cMpc.  Lengths of lines indicate orientations
  with respect to the projection, long lines are nearly perpendicular
  to the line of sight and short lines are nearly parallel to the line of sight, respectively.  
  Grey scales correspond to the surface mass
  density of DM component which are computed by the projection of all particles
  in the simulation box at middle time for each panel
  $t=1.97$ (top), $t=0.67$ (middle), $t=0.16$ (bottom) Gyr, respectively.
  }
    \label{fig:orientations}
\end{figure*}

This paper has examined the correlation of orientations of the
central galaxies (CGs) and their host dark matter (DM) haloes extracted from the
Horizon-AGN cosmological hydrodynamical simulation \citep{dubois14}.
We identified 40 cluster-sized DM haloes at $z\approx0$, and
traced their progenitor haloes and CGs at 50 different redshifts up to $z\sim5$. 
By applying the three-dimensional ellipsoidal fitting to
those objects, we adopted the direction of their major axes ${\bm
  \hat{a}}_1(t)$ as a measure representing their orientations.  In
addition, we computed the eigen-vectors of the tidal field centred at
the location of the CG in each halo, and found that ${\bm
  \hat{u}}_3(t)$ smoothed over $10~h^{-1}$cMpc corresponding to the
direction of the {\it slowest collapsing} (or even stretching) mode is
a good proxy characterising the orientation of the large-scale
structure surrounding each object.

A picture of the evolution of the orientations of CGs and DM haloes
emerging from our current study is summarized as follows.
Even at early epochs ($t<4$ Gyr), orientations of the CG and its host DM
halo in an individual system exhibit significant correlation in a statistical sense.
The orientations of the CG and host DM halo
are well aligned at each epoch, and their alignment becomes tighter toward the present epoch.
On the other hand, the orientations of both the CG and its host DM halo significantly change
due to mergers and continuous mass accretions;
the orientations of the CG and host DM halo change
coherently, and evolve together toward their current orientations
that are more tightly correlated with the surrounding large-scale matter distribution ${\bm \hat{u}}_3(t_0)$ than at early epochs.
This implies that the instantaneous alignment between the DM halo and the CG is driven 
by strong dynamical interactions through repeated mergers and mass accretions.
Since the direction of ${\bm \hat{u}}_3(t)$
barely changes over the cosmic time, 
the current orientation of the DM halo, and
therefore that of the CG, is basically imprinted in the primordial
density field of the Universe. Indeed the CG evolves following that
of the host DM halo and becomes tightly aligned with each other;
their typical angles are $< 30^\circ$ and $< 20^\circ$ in the three
dimensional space and in the projected plane, respectively, at the present epoch.

The above basic picture is visually illustrated in Figure
\ref{fig:orientations}. Each panel depicts the simulation box of
$(100~h^{-1}{\rm cMpc})^3$ projected along the $z$-axis of the
simulation. The grey scale represents the surface density of DM component on
$(1~h^{-1}{\rm cMpc})^2$ cells at $z=1.97$ (top), 0.67 (centre)
and 0.16 (bottom). Green bars in the left panels and red bars in the
right panels indicate the eigen-vector ${\bm\hat{u}}_3(t)$ of the
tidal field and the major axis ${\bm \hat{a}}_1(t)$ of CGs projected
on each $x$-$y$ plane, whereas blue bars in all the panels are the
projected major axis ${\bm \hat{a}}_1(t)$ of DM haloes at epochs around
the redshift of each panel.
The green bars are roughly aligned along the filamentary structure and
do not change so much. The blue bars seem to be aligned with the green
bars gradually with time, and the tendency of the mutual alignment is
stronger between the blue and red bars, i.e., DM haloes and CGs.

In this paper, we presented the predicted evolution of alignments between BCGs, DM haloes, 
and the large-scale structure, which should be confronted with observations.
A caveat is that we focused on the evolution of the same halo 
over the cosmic time whose mass is different at different epochs (see Figure \ref{fig:t_ar_no2}).
Such difference of masses should be taken
into account for a fair comparison with observations \citep{2017ApJ...851..139L}.
The survey result by Hyper Suprime-Cam Survey \citep{2018PASJ...70S...8A} would be useful for examining the
redshift evolution of the alignment between orientations of BCGs and
clusters because it covers a large ($\sim1000$ deg$^2$) and deep ($z\sim1.1$) area \citep{2018PASJ...70S..20O}.

\section*{Acknowledgements}
We thank J. Devriendt and the Horizon-AGN team for making their
simulation data available to us.  We thank N. E. Chisari for sharing
the draft of \cite{bate2019} with us and also for many useful
comments.  We also thank S. Codis for useful comments.  T.O. is
supported by Advanced Leading Graduate Course for Photon Science
(ALPS) at the University of Tokyo.  This work is supported partly by
Japan Society for the Promotion of Science (JSPS) Core-to-Core Program
“International Network of Planetary Sciences”, and also by JSPS
KAKENHI Grant Numbers JP17J05056 (T.O.), JP17K14273 (T.N.), JP18K03693
(M.O.), JP15H05892 (M.O.), JP18K03704 (T.K.), and JP18H01247 and
JP19H01947 (Y. S.).  T.N. acknowledges Japan Science and Technology
Agency (JST) CREST Grant Number JPMJCR1414.


\bibliographystyle{mnras}
\bibliography{reference} 

\appendix

\bsp	
\label{lastpage}
\end{document}